\documentclass[aps,twocolumn,prb]{revtex4}

\usepackage{graphicx}
\def\be{\begin{equation}}
\def\ee{\end{equation}}

\def\ie{{\it i.e.}}

\def\t1{$T_1^{-1}$}
\def\iTtwo{$T_2^{-1}$}
\def\iT2{$T_2^{-1}$}
\def\se{$^{77}$Se}

\def\cl{(TMTSF)$_2$ClO$_4$}

\def\pf{(TMTSF)$_2$PF$_6$}

\def\a{$\mathbf{a}$}

\def\cs{$\mathbf{c^{\ast}}$}

\begin{document}

\title{NMR evidence for very slow carrier density fluctuations in the organic 
metal (TMTSF)$_2$ClO$_4$}

\author{F. Zhang}
\email[]{fzhang@physics.ucla.edu}
\author{Y. Kurosaki}
\author{J. Shinagawa}
\author{B. Alavi}
\author{S. E. Brown}
\affiliation{Department of Physics $\&$ Astronomy, UCLA, Los Angeles, California
90095}

\voffset=1cm
\begin{abstract}
We have investigated the origin of the large increase in spin-echo decay rates
for the $^{77}$Se nuclear spins at temperatures near to $T=30K$ in the organic
superconductor (TMTSF)$_2$ClO$_4$. The measured angular dependence of $T_2^{-1}$
demonstrates that the source of the spin-echo decays lies with carrier density 
fluctuations rather than fluctuations in TMTSF molecular orientation. The very 
long time scales are directly associated with the dynamics of the anion ordering
occurring at $T=25K$, and the inhomogeneously broadened spectra at lower 
temperatures result from finite domain sizes. Our results are similar to 
observations of line-broadening effects associated with charge-ordering 
transitions in quasi-two dimensional organic conductors.

\end{abstract}

\pacs{74.70.Kn, 75.30.Gw, 76.60.-k}

\maketitle

\cl\ is a member of the Bechgaard salts family of organic conductors and
superconductors. While the original discovery of organic superconductivity was
made in the isostructural system \pf\ at Orsay \cite{Jerome1980}, the
perchlorate salt was interesting because it was the only composition undergoing
a superconducting transition ($T_c=1.4K$) at ambient pressure \cite{Parkin1981,
Ishiguro1998}. Another difference, when compared with the hexaflouride salts is
that the counterion is non-centrosymmetric. As a result, there is a symmetry
breaking of the high-temperature crystallographic space group (Pmma), when the
ClO$_4$ counterions orientationally order at $T_{AO}=25K$ with wavevector
$\mathbf{Q}=(0,1/2,0)$ \cite{Pouget1996}. This particular $\mathbf{Q}$ leads to
inequivalent TMTSF stacks and therefore also two bands crossing the Fermi energy
$E_F$. Some of the low-temperature properties of \cl\ that differ from the
PF$_6$ salt are attributed to this difference \cite{Lebed2004,Kang1993,
Chashechkina1997,Chashechkina1998}.

The phase transition at $T_{AO}$ is a metal-metal phase transition driven by
lattice Coulomb effects at which a disproportionation, or charge order (CO), of
carrier density between the adjacent stacks accompanies the counterion ordering.
By now, CO is known to be ubiquitous to the analog TMTTF-based salts with
counterions ranging from hexaflourides like AsF$_6$ to non-centrosymmetric SCN
or ReO$_4$. Those systems are all insulators, either resulting from the
broken-symmetry of the CO phase transition itself \cite{Seo1997,Chow2000,
Monceau2001} or resulting from a dimerization of intrastack bond distances
between TMTTF molecules \cite{Emery1982,Jerome2002}. In the case of the
insulators, there is little doubt that electron-electron interactions play a
crucial role; nevertheless there is empirical evidence that coupling to the
counterions allows the CO transition to take place. We examine the effects of
the transition on the NMR properties as a window into what differences might be
observed when compared to what is seen in the insulators.

Here we explore the effects of the anion ordering on the \se\ NMR spectrum and
relaxation in \cl. In this case, the transverse spin relaxation rate is strongly
affected by slow fluctuations of the electronic carrier density above $T_{AO}$. 
The slow fluctuations have been observed previously by Takigawa and Saito (TS) 
\cite{Takigawa1986}, but with the difference that our experiments are done using
single crystals, and we are able to demonstrate that the carrier density 
fluctuations rather than molecular orientational fluctuations are responsible. 
The linewidth broadens homogeneously in association with these fluctuations and 
narrows on cooling further. From experiments undertaken below $T_{AO}$ on 
quenched and relaxed samples, we conclude that the spectrum is dominated by 
disorder effects. This is very different from the insulators, where distinct 
symmetry-breaking signatures of the ordered phase are observed 
\cite{Zamborszky2002}. Even in the presence of the disorder, \cl\ is a 
superconductor, and the pairing is probably not s-wave.

The \cl\ crystals were grown using standard electrolysis techniques.
In this case, the dimensions are $6.4mm \times 1.2mm \times 0.8mm$
and the mass is $m=4.2 mg$. Typically, crystals grow longer along
the highly-conducting stack direction relative to the other
directions. Our goal was to investigate the effect on the hyperfine
fields resulting from the ClO$_4$ anion ordering, so the appropriate
coil geometry is unusual when compared to most NMR investigations of
the Bechgaard salts. We expected the hyperfine fields are
principally dipolar \cite{Takigawa1986}, arising from the $p_z$
orbital on the Se atoms, extending out of the plane of the molecule.
As a result, the strongest hyperfine fields are with the dc field
aligned along the stacking axis, so we chose the symmetry axis of
the coil to allow for sample rotations with the external field 
$\mathbf{B_0}$ lying in the \a -\cs\ plane. The AO
transition occurs at $T_{AO}=25K$, so we cooled at $40mK/min.$
through it to reach the highly-conducting relaxed state.

\begin{figure}
\includegraphics[width=3.2in]{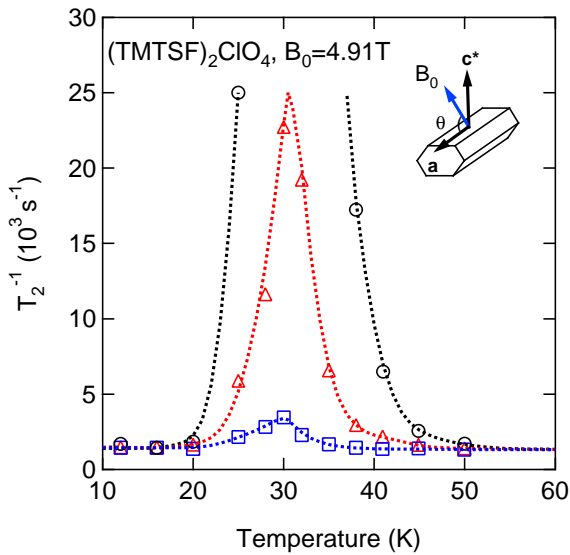}
\caption{(Color Online) Temperature dependence of \iT2\ for three 
orientations of the crystal relative to the external magnetic field. 
The inset illustrates the geometry. $\theta$ 
indicates the angle between the direction of applied field $\mathbf{B_0}$ 
and \a\ axis. The open circles are taken with 
$\mathbf{B_0}\parallel$ \a\ ($\theta=0$), the open triangles are taken with 
$\mathbf{B_0}\parallel$ \cs\ ($\theta=90^{\circ}$), and the open squares are 
taken with $\theta=\theta_m$. The dotted lines serve as guides to the eyes.} 
\label{iT2vTmp}
\end{figure}

Shown in Fig. \ref{iT2vTmp} is the spin-echo decay rate $T_2^{-1}$
(here we define $T_2$ as the time corresponding to $1/e$ decay) {\it{vs.}}
temperature for several angles of the applied field $B_0=4.91T$. As
observed by Takigawa and Saito \cite{Takigawa1986}, we see a
substantial increase in the relaxation rate on lowering the
temperature through $30K$. The questions we answer are how this
increase in $T_2^{-1}$ is related to anion ordering and the
mechanism by which it occurs. By rotating the sample, we establish 
the anisotropy in \iT2\ in the vicinity of $T=30K$. The size of the 
peak is found largest for the field applied
along the stacking axis $(\mathbf{B_0}\parallel\mathbf{a}, \theta=0)$,
somewhat smaller for the \cs-direction, and even less when $\mathbf{B_0}$ 
is rotated away from $\mathbf{a}$ by $\theta=\theta_m\equiv55^\circ$ 
(inset; also see Eq. \ref{shiftvA} below, noting that 
3cos$^2\theta_m-1\equiv0$), where the anisotropic part of the hyperfine 
field vanishes. At the peak, the rate is too fast to allow for a 
measurement with the field along \a.

\begin{figure}
\includegraphics[width=3.2in]{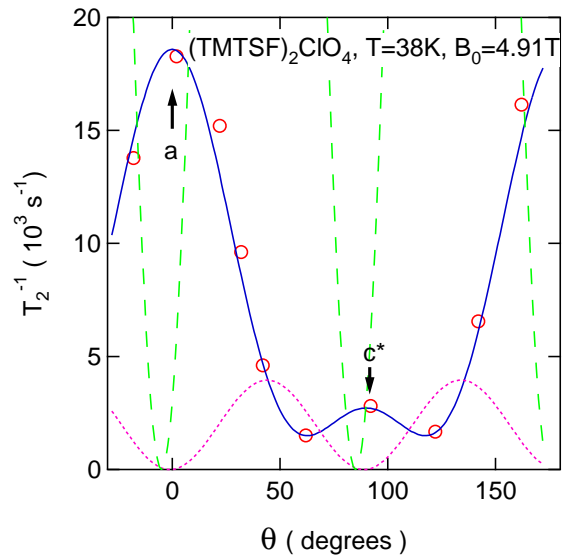}
\caption{(Color Online) Angular dependence of 
$T_2^{-1}$ (red open circles) at $T=38K$. The pink dotted
line and the green dashed line are the expected angular dependence of
$T_2^{-1}$ in TS model, corresponding to the molecular motion
amplitudes of $2^{\circ}$ and $10^{\circ}$, respectively.}
\label{iT2vTheta}
\end{figure}

A more complete angular dependence of the spin-echo decay rate is shown in
Fig. \ref{iT2vTheta} at $T=38K$, where the relaxation rate is
enhanced over a temperature-independent background. We also notice
that there is a change in linewidth and lineshape on cooling through
the temperatures where $T_2^{-1}$ peaks. In the high temperature
regime (\ie, $T\approx50K$), the lineshape is very close to
Gaussian. It changes to Lorentzian on the high-temperature side of
the peak and returns to Gaussian (although considerably broader) on
the low-temperature side of the peak. 

In what follows, we make the case that the peak in \iTtwo\ {\it{vs.}} $T$
is a result of the slowing down of carrier density fluctuations 
linked to the inequivalence of TMTSF stacks developing at the anion
ordering transition. The spin-echo decay is affected by the density
fluctuations through the z component of the hyperfine fields,
because local temporal variations result in dephasing of the
precessing spins involved in a spin-echo experiment. To be more
specific, suppose that a nuclear spin is situated in an environment
in which the local field switches randomly between two discrete
$\pm\delta h_z$ on a time scale $\tau_c$. If this is the only source
for the spin-echo decay, then the spin-echo amplitude decreases as 
\be 
G(t)=G(0)e^{-(\gamma \delta h_z)^2\tau_c t}, 
\label{motionaleqn}
\ee 
in the extreme narrowing limit ($\gamma \delta h_z\tau_c\ll 1$), with 
$\gamma$ the gyromagnetic ratio. When Eq. \ref{motionaleqn} applies, the 
spectrum is said to be motionally narrowed. The spin-echo decay rate
increases when $\gamma \delta h_z\tau_c$ grows larger, as it would on the
approach to the anion ordering transition. Dephasing is most efficient when 
$\gamma \delta h_z\tau_c\sim1$, where we expect a peak in \iT2.  
Below, we evaluate the form of the hyperfine coupling, and use it to 
describe the angular dependence of the spin-echo decay rate.

\begin{figure}
\includegraphics[width=3.2in]{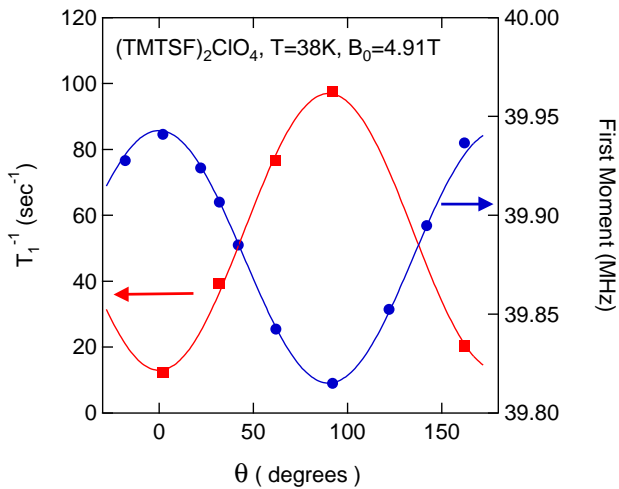}
\caption{(Color Online) First moment of the spectra (solid circles) and
spin-lattice relaxation rate $T_1^{-1}$ (solid squares) {\it{vs.}} angle 
$\theta$ at $T=38K$.}
\label{iT1vTheta}
\end{figure}

The angular dependence of the first moment as well as the spin-lattice
relaxation rate $T_1^{-1} $at $T=38K$ are shown in Fig.
\ref{iT1vTheta}. We assume these are consistent with a uniaxial
hyperfine coupling of the form
\begin{eqnarray}
A(\theta) &=& A_{iso}+A_{ax}(3\cos^2\theta-1),\label{shiftvA}\\
K(\theta) &=& A(\theta)\chi^s\frac{v}{2},
\label{shifteqn}
\end{eqnarray}
with $v$ the volume per formula unit and $\chi^s$ the dimensionless
spin susceptibility. $A_{iso}$ is the contribution from core
polarization of the Se ions. We assume that $A_{ax}$ is dominated
by the intra-atomic Se $p_z$ orbital and takes the form
$A_{ax}=(2/5)<r^{-3}>\sigma$ with $\sigma$ the carrier density on a
single ion. The upper limit for $\sigma$ is 0.25; it is reasonable
to expect less, and TS used $\sigma=0.19$ calculated in Ref.
\cite{Metzger1981}. The stacking (\ie, \a) axis is very nearly
coincident with the principal axis orthogonal to the plane of the
TMTSF molecule.

Then the hyperfine field contribution to the spin-lattice relaxation
rate also varies with angle, even though the spin susceptibility is
isotropic \cite{Miljak1988}. For $\mathbf{B_0}\parallel\mathbf{a}$
($\theta=0$) and where correlations are ignored, we have
\begin{equation}
\frac{1}{T_1^aT}=4\pi\frac{k_B}{\hbar}\left(\frac{^{77}\gamma}{\gamma_e}
\right)^2\left[(A_{iso}-A_{ax})\chi^s\frac{v}{2}\right]^2,
\label{iT1eqn}
\end{equation}
with $\gamma_e$ the gyromagnetic ratio of the free electron. For the angular
dependence of the relaxation rate, we obtain
\begin{equation}
\frac{T_1^a}{T_1(\theta)}=\frac{(\alpha+2)^2\sin^2\theta
+(\alpha-1)^2(2-\sin^2\theta)}{2(\alpha-1)^2},
\label{iT1Aeqn}
\end{equation}
where $\alpha=A_{iso}/A_{ax}$.

Inserting the accepted value for the spin susceptibility of
$\chi^s=1.6\times10^{-4} emu/mole (f.u.)$ \cite{Miljak1988} into
Eq. \ref{shifteqn}, together with $K_{ax}=10.7\times10^{-4}$ extracted from
the angular dependence of the first moment (Fig. \ref{iT1vTheta}), we obtain a 
value for $<r^{-3}>=15a_0^{-3}$, where $a_0$ is the Bohr radius. This value 
is larger than the $9.3a_0^{-3}$ obtained for Se atoms from Hartree-Fock 
calculations \cite{Fraga1976}. The solid line running through the relaxation 
rate data shown in Fig. \ref{iT1vTheta} is a least square fit to 
Eq. \ref{iT1Aeqn}, from which we extract $\alpha=0.37$ \cite{SLRRnote}. 
Returning to Fig. \ref{iT2vTheta} and using this value of $\alpha$, we 
obtain the solid line through the data points using the function
\begin{equation}
T_2^{-1}=\beta^2(K(\theta)\gamma B_0)^2\tau_c(T) + B^2,
\label{iT2vAeqn}
\end{equation}
with $B_0$ the external applied magnetic field, and $B$ a
temperature and angle independent constant. The first term of 
Eq. \ref{iT2vAeqn} results from local field fluctuations originating in 
variations of the hyperfine coupling (Eq. \ref{shifteqn}), which are 
caused by carrier density fluctuations. At higher temperatures,
$\tau_c$ is sufficiently short so as to make hyperfine field
fluctuations inconsequential for the spin-echo decay, and it grows upon
cooling with the slowing of the lattice fluctuations. $\beta$ is a
constant that measures the amplitude of the hyperfine field
fluctuations relative to the average.

Although the dependence of the spin-echo decay rate on the external field
is the same as in TS model, the dependence on angle is very different to
that expected for orientational fluctuations of the molecule. For
contrast, we show the relative angular dependence of this mechanism
for orientational fluctuation amplitudes of $\delta\theta=2^{\circ}, 
10^{\circ}$ in Fig. \ref{iT2vTheta}. In that case, the spin-echo decay rates
induced by the local field fluctuations are largest where
$dK/d\theta$ is maximum, namely near to $\theta_m$.
Instead, we observe a minimum at $\theta_m$. The angular dependence is
consistent with fluctuations in the hyperfine coupling, which
includes the factor $\sigma$ representing the fractional number of carriers 
per Se atom. Our interpretation is that as the correlation length for the
counterion ordering grows, so does the characteristic time scale for
the fluctuations in the usual way. In the ordered state there is an
inequivalent carrier density on adjacent stacks, as a result of the
period doubling of the lattice in the $\mathbf{b}$ direction. The
fluctuations in the component of the hyperfine field parallel to the
applied field follow. What is striking is that the time scales are
governed by the lattice motion, while the hyperfine fields merely
probe it. 

\begin{figure}
\includegraphics[width=3.2in]{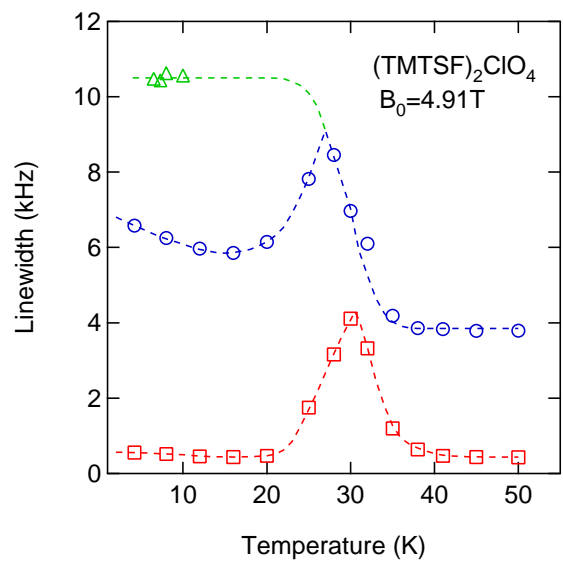}
\caption{(Color Online) Homogeneous and inhomogeneous linewidth as a function of
temperature, taken with $\mathbf{B_0}\parallel$ \cs. The open circles are the
inhomogeneous linewidth of the spectra, defined as the square root of the second
moment. The open squares are the homogeneous linewidth, defined as the HWHM of 
the Fourier Transformation of the spin-echo decay. The triangles are the 
linewidth taken in the completely quenched state.
The lines through the data points are guides to the eyes.}
\label{LWvTc}
\end{figure}

The temperature dependence of the homogeneous and inhomogeneous linewidth 
for $\mathbf{B_0}\parallel\mathbf{c^{\ast}}$ is shown in Fig. \ref{LWvTc},
both in the relaxed and completely quenched state. At high temperatures,
the spectra are dominated by the inhomogeneous broadening due to the 
inequivalent Se sites. The line is then homogeneously broadened on the high
temperature side of the \iT2\ peak and is narrowed again on the low 
temperature side of the peak in the relaxed state. At lower temperatures 
(below $10K$), the inhomogeneous broadening grows significantly without 
evidence for saturation.

The fluctuation amplitude of the hyperfine field can be
estimated from the low temperature linewidth in the completely
quenched state (Fig. \ref{LWvTc}) from which we obtain $\gamma \delta h_z/2\pi
=O(10^4)Hz$. Using this value of $\delta h_z$ and $K(\theta=90^{\circ})
=6.7\times10^{-4}$, $\beta$ can be evaluated as $0.4$. We can also estimate the 
characteristic time scale of the fluctuations near the peak to be 
$\tau_c=O(\frac{1}{2\pi}\times 10^{-4})s$. On the low temperature side of the 
peak of \iT2\, the fluctuations are so slow as to be ineffective in dephasing 
the transverse nuclear spin components in a spin-echo experiment. That we should 
observe this trend above $T_{AO}$ should not be surprising because only on
very slow cooling through the anion ordering transition results in a
relaxed phase and superconductivity.

It remains to explain the nature of the low temperature inhomogeneous 
broadening in Fig. \ref{LWvTc}, which has a similar angular dependence as 
the homogeneous broadening peaking near $30K$. Instead of two discrete 
peaks in the spectrum, there is a single broad feature. As a result of 
the anion ordering, there is a distinction between adjacent chains, say 
``A'' chains and ``B'' chains. Zone folding produces two bands crossing 
the Fermi energy, where the degeneracy at the zone face near
$(\pm k_F,\pm\pi/2b,q_c)$ is lifted by a small gap, $\Delta_{AO}$. 
There is a corresponding difference in Fermi wavevector,
\be
\Delta k_F=\frac{\Delta_{AO}}{\hbar v_F}.
\ee
The two bands are of predominantly ``A'' or ``B'' character for
these states near to the zone edge, but they are mixed if $\xi\Delta
k_F\le2\pi$, with $\xi$ the correlation length of the counterion
ordering measured along the stacks. Using $\Delta_{AO}\approx 2meV$
\cite{Uji1996,Lebed2004} and $v_F=2\times10^5m/s$ gives $\Delta
k_F=O(10^{-3})$\AA$^{-1}$. Therefore, quite large values of $\xi$ are
necessary to maintain the distinction between ``A'' and ``B'' chains.
Perhaps sufficient order for observation of distinct chains could be
achieved with slower cooling. Indeed, in magnetotransport studies
\cite{Uji1996}, much slower rates ($\sim10mK/min.$) were necessary
to observe interference effects linked to the existence of the gap
induced by the anion ordering \cite{Uji1996}.

If we presume the growth of the inhomogeneous broadening at low temperatures
(Fig. \ref{LWvTc}) is linked to the configuration of the counterions, 
then either it is a dynamic effect linked to motional narrowing associated 
with lattice defect mobility, or it is a static effect, where the nature
of the defects in the counterion sublattice changes as the
temperature is lowered. For the latter, an example is where the
domain boundary between ``A'' and ``B'' stacks broadens upon
cooling, perhaps as a consequence of thermal contraction of the
lattice. Experiments sensitive to spectral diffusion could
distinguish between the two mechanisms for the broadening.

Finally, we make some general comments on our observations in the context of
charge ordering in organic conductors. In our view, the reason for the very slow
fluctuations evident here are the large masses and moments of inertia involved
in the anion ordering transition. As a result of the coupling of the electronic
states to the lattice potential, this becomes apparent in the spin-echo decay 
and inhomogeneous line broadening at low temperatures. Very similar results for 
spin-echo decays and inhomogeneous broadening were recently reported in $^{13}$C
NMR experiments on $\theta$-(BEDT-TTF)$_2$RbZn(SCN)$_4$ \cite{Chiba2004}, a
quasi-two dimensional organic conductor. In that case, there is strong evidence
that the slow fluctuations are directly associated with a charge-ordering phase
transition, also linked to metal-insulator transition, for the 1/4-filled
system. We suggest that the slow fluctuations are evidence that coupling to the
lattice is an important component of the CO transition in that system, and
perhaps all of the quasi-two dimensional organic systems undergoing charge order
symmetry breaking.

This work was supported by the National Science Foundation under grant
No. DMR-0203806. We acknowledge helpful discussions with Serguei Brazovskii.

%\bibliography{ClO4}

\end{document}